\documentclass[11pt]{article}

\usepackage{amsfonts,amssymb,chet,dsfont,graphicx,mathrsfs,pgfplots,tikz}

\usepackage{subcaption}
\usetikzlibrary{external}
\title{A Probability Density Function for\\Neutrino Masses and Mixings}

\author{Jean-Fran\c{c}ois Fortin\email{jean-francois.fortin@phy.ulaval.ca}, Nicolas Giasson\email{nicolas.giasson.1@ulaval.ca} and Luc Marleau\email{luc.marleau@phy.ulaval.ca}}

\affiliation{
D\'epartement de Physique, de G\'enie Physique et d'Optique,\\Universit\'e Laval, Qu\'ebec, QC G1V 0A6, Canada
}

\abstract{The anarchy principle leading to the see-saw ensemble is studied analytically with the usual tools of random matrix theory.  The probability density function for the see-saw ensemble of $N\times N$ matrices is obtained in terms of a multidimensional integral.  This integral involves all light neutrino masses, leading to a complicated probability density function.  It is shown that the probability density function for the neutrino mixing angles and phases is the appropriate Haar measure.  The decoupling of the light neutrino masses and neutrino mixings implies no correlation between the neutrino mass eigenstates and the neutrino mixing matrix and leads to a loss of predictive power when comparing with observations.  This decoupling is in agreement with some of the claims found in the literature.}

\date{September 2016} 

\begin{document}

\maketitle



\section{Introduction}\label{SIntro}

The Standard Model (SM) is the pinnacle of our understanding of particle physics.  It however comes with a plethora of parameters, the masses and the flavor mixings, that are seemingly not fixed by any known fundamental principle.  On one hand, the spectrum of quarks spans five orders of magnitude and the quark mixing matrix exhibits some form of flavor hierarchy.  On the other hand, the spectrum of charged leptons spans four orders of magnitude while the neutrinos are strictly massless, hence the lepton mixing matrix is trivial.  Observations of neutrino oscillations \cite{pdg,Gonzalez-Garcia:2014bfa} however cannot be accommodated in the SM and this suggests to extend the SM to allow for neutrino masses and neutrino mixings.  Experimental data show that the neutrino sector prefers three massive light neutrinos with different masses and a neutrino mixing matrix exhibiting near-maximal mixing.

To make sense of the neutrino sector, it was argued in \cite{Hall:1999sn,Haba:2000be} that the light neutrino mass matrix could be generated randomly from a more fundamental Dirac neutrino mass matrix and a more fundamental Majorana neutrino mass matrix with random elements distributed according to a Gaussian ensemble, a principle dubbed the anarchy principle.  These more fundamental neutrino mass matrices would come from the extended SM where the see-saw mechanism occurs.  In \cite{Haba:2000be}, it was argued that the probability density function (pdf) for the mixing angles and phases is the appropriate Haar measure of the symmetry group, implying near-maximal mixings.  Then, the anarchy principle was analysed mostly numerically in a series of papers \cite{deGouvea:2003xe,deGouvea:2012ac,Heeck:2012fw,Bai:2012zn,Lu:2014cla}, reaching interesting conclusions, for example about the preferred normal hierarchy of the neutrino masses.

Although several numerical results have been obtained, few analytical results on the see-saw ensemble, which is derived from the anarchy principle, exist.  That it is the case even though random matrix theory is a well-studied subject in mathematics \cite{mehta2004random,muirhead2009aspects,forrester2010log} is surprising.  It is therefore clear that a thorough analytical investigation of the see-saw ensemble is possible.

This paper undertakes this task by investigating analytically the see-saw ensemble derived from the anarchy principle with the help of the usual tools of random matrix theory.  The see-saw ensemble pdf is obtained from $N\times N$ fundamental Dirac and Majorana neutrino mass matrices with real or complex elements.  The result is given in terms of a specific multidimensional integral.  The joint pdf for the singular (eigen) values in the complex (real) case is then derived and it is shown that the group variables decouple straightforwardly as in the usual Gaussian ensembles.  Simple properties of the see-saw ensemble pdf are also presented and their implications for the physical case of $N=3$ are briefly discussed.  The full investigation of the see-saw ensemble for the physical case of interest, SM neutrino physics with $N=3$, will be discussed elsewhere \cite{Fortin:2017iiw}.

This paper is organized as follows: Section \ref{STypeISS} quickly reviews the type I see-saw mechanism.  In section \ref{SSeesaw} the pdf for the see-saw ensemble is obtained from the pdfs related to the Dirac and Majorana neutrino mass matrices.  Section \ref{SProperties} gives the general properties of the see-saw ensemble from the pdf obtained previously.  A discussion, including a comparison with the existing literature, and a conclusion are presented in section \ref{SConclusion}.  Finally, appendices \ref{SMeasures} and \ref{SCircle} derive explicitly some useful results.


\section{Type I See-Saw Mechanism}\label{STypeISS}

This section reviews the type I see-saw mechanism \cite{Minkowski:1977sc}.

\subsection{Type I See-Saw Mechanism}\label{SsTypeISS}

In the SM, neutrinos, which are part of the doublets $L^i$ where $i$ denotes one of the three families, are massless.  To account for the oscillations observed experimentally, the type I see-saw mechanism postulates the existence of three singlet fermions $N^i$ (left-handed Weyl spinors) that play the role of the ``missing'' right-handed neutrinos $\sigma^2N^{i*}$.  In terms of the SM Lagrangian, the type I theory is described by
\eqn{\mathcal{L}=\mathcal{L}_\text{SM}+\bar{N}^ii\bar{\sigma}^\mu\partial_\mu N^i-\left(\tfrac{1}{2}M_{R,ij}N^iN^j+y_{D,ij}L^iHN^j+\text{h.c.}\right),}[EqnLI]
where $H$ is the Higgs doublet and the allowed terms correspond to right-handed neutrino masses $M_R^{ij}$ and Yukawa interactions $y_D^{ij}$.  From \eqref{EqnLI} the $(\nu\quad N)$ neutrino mass matrix is given by
\eqn{M_{\nu N}=\left(\begin{array}{cc}0&M_D\\M_D^T&M_R\end{array}\right),}[EqnMI]
where $M_D=y_Dv/\sqrt{2}$ is the Dirac neutrino mass matrix and $v$ is the Higgs vacuum expectation value, \textit{i.e.} $\langle H\rangle=\frac{1}{\sqrt{2}}(0\quad v)^T$.

The right-handed neutrino mass scale could naturally be much larger than the Dirac neutrino mass scale since the former is an (\textit{a priori} arbitrary) energy scale while the latter originates from a Yukawa coupling, hence \eqref{EqnMI} implies that the light neutrino mass matrix is
\eqn{M_\nu=-M_DM_R^{-1}M_D^T.}[EqnMnu]
The (symmetric) neutrino mass matrix \eqref{EqnMnu} is the starting point to compute the see-saw ensemble pdf from the pdfs for the Dirac and Majorana neutrino mass matrices $M_D$ and $M_R$, as described in the next section.


\section{Definition of the See-Saw Ensemble}\label{SSeesaw}

In this section we obtain the pdf for the type I see-saw ensemble from the pdfs associated to the Dirac and Majorana neutrino mass matrices.  To be as general as possible, the fundamental neutrino mass matrices are taken to be $N\times N$ instead of $3\times3$.

\subsection{Dirac and Majorana Ensembles}\label{SsDM}

Following the principle of anarchy \cite{Haba:2000be,Lu:2014cla}, where elements of the fundamental matrices $M_D$ and $M_R$ are random, the pdfs for the Dirac and Majorana neutrino mass matrices are Gaussian and defined as
\eqna{
P_D(M_D)dM_D&=\left(\frac{1}{2\pi\Lambda_D^2}\right)^{\beta N^2/2}\exp\left[-\frac{\text{tr}(M_D^\dagger M_D)}{2\Lambda_D^2}\right]dM_D,\\
P_R(M_R)dM_R&=\left(\frac{1}{\pi\Lambda_R^2}\right)^{\beta N(N-1)/4}\left(\frac{1}{2\pi\Lambda_R^2}\right)^{\beta N/2}\exp\left[-\frac{\text{tr}(M_R^\dagger M_R)}{2\Lambda_R^2}\right]dM_R,
}[EqnPDFdM]
where $\beta=1$ corresponds to real matrix elements and $\beta=2$ corresponds to complex matrix elements.  The difference between the two pdfs comes from the fact that $M_R$ is symmetric while $M_D$ is not.  The pdfs \eqref{EqnPDFdM} are properly normalized with respective standard deviations $\Lambda_D$ and $\Lambda_R$ (for the diagonal elements only, the off-diagonal elements have standard deviations $\Lambda_R/\sqrt{2}$).

Before proceeding, it is convenient to re-express the pdfs in terms of the dimensionless variables $\hat{M}_D=M_D/(\sqrt{2}\Lambda_D)$ and $\hat{M}_R=M_R/(\sqrt{2}\Lambda_R)$.  Hence the pdfs \eqref{EqnPDFdM} become
\eqna{
P_D(\hat{M}_D)d\hat{M}_D&=\tilde{C}_{DN}^\beta e^{-\text{tr}(\hat{M}_D^\dagger\hat{M}_D)}d\hat{M}_D,\\
P_R(\hat{M}_R)d\hat{M}_R&=\tilde{C}_{RN}^\beta e^{-\text{tr}(\hat{M}_R^\dagger\hat{M}_R)}d\hat{M}_R,
}[EqnPDFdMh]
where the normalization constants are
\eqn{\tilde{C}_{DN}^\beta=\frac{1}{\pi^{\beta N^2/2}}\quad\quad\text{and}\quad\quad\tilde{C}_{RN}^\beta=\frac{2^{\beta N(N-1)/4}}{\pi^{\beta N(N+1)/4}}.}[EqntC]

By the singular (eigen) value decomposition theorem, both matrices can be diagonalized with non-negative (real) elements as $\hat{M}_D=U_LD_DU_R^\dagger$ and $\hat{M}_R=UD_RU^T$ where the $U$-matrices are unitary (orthogonal) for $\beta=2$ ($\beta=1$).  In terms of the singular values and the remaining variables, the measure for the Dirac neutrino mass matrix can be simplified to
\eqn{d\hat{M}_D=c_{AN}^\beta\prod_{1\leq i<j\leq N}|\hat{m}_{D,i}^2-\hat{m}_{D,j}^2|^\beta\prod_{1\leq i\leq N}\hat{m}_{D,i}^{\beta-1}d\hat{m}_{D,i}U_{LR}^\dagger dU_{LR},}[EqndMD]
where the measure $d\hat{M}_D'=d\hat{M}_D$ is invariant under a transformation $\hat{M}_D'=U_L\hat{M}_DU_R^\dagger$ and $U_{LR}^\dagger dU_{LR}$ is the appropriate group (Haar) measure.  Moreover, the measure for the right-handed neutrino mass matrix, which is invariant under $\hat{M}_R'=U\hat{M}_RU^T$, \textit{i.e.} $d\hat{M}_R'=d\hat{M}_R$, can be written as
\eqn{d\hat{M}_R=c_{SN}^\beta\prod_{1\leq i<j\leq N}|\hat{m}_{R,i}^\beta-\hat{m}_{R,j}^\beta|\prod_{1\leq i\leq N}|\hat{m}_{R,i}|^{\beta-1}d\hat{m}_{R,i}U^\dagger dU,}[EqndMR]
where $U^\dagger dU$ is again the appropriate group (Haar) measure.  The proofs of \eqref{EqndMD} and \eqref{EqndMR} are given in appendix \ref{SMeasures}.  The normalization constants for arbitrary and symmetric matrices $c_{AN}^\beta$ and $c_{SN}^\beta$ can also be found in the appendix and are given by \eqref{EqnAconstant} and \eqref{EqnSconstant} respectively.  The absolute value is not necessary for the Dirac matrix because all singular values are non-negative.  On the other hand, the absolute value is absolutely necessary for the real Majorana matrix ($\beta=1$) because the decomposition in this case is an eigenvalue decomposition where all eigenvalues can also be negative.  In the following, the expression ``singular values'' will be used for both singular values and eigenvalues.

Since $P_D(\hat{M}_D)$ and $P_R(\hat{M}_R)$ depend only on the singular values, it is easy to write the joint pdfs for the singular values,
\eqna{
P_D(\hat{m}_D)d\hat{m}_D&=C_{DN}^\beta\prod_{1\leq i<j\leq N}|\hat{m}_{D,i}^2-\hat{m}_{D,j}^2|^\beta\prod_{1\leq i\leq N}\hat{m}_{D,i}^{\beta-1}e^{-\hat{m}_{D,i}^2}d\hat{m}_{D,i},\\
P_R(\hat{m}_R)d\hat{m}_R&=C_{RN}^\beta\prod_{1\leq i<j\leq N}|\hat{m}_{R,i}^\beta-\hat{m}_{R,j}^\beta|\prod_{1\leq i\leq N}|\hat{m}_{R,i}|^{\beta-1}e^{-\hat{m}_{R,i}^2}d\hat{m}_{R,i},
}[EqnPDFdmh]
where the normalization constants are
\eqna{
C_{DN}^\beta&=\tilde{C}_{DN}^\beta c_{AN}^\beta\int U_{LR}^\dagger dU_{LR}=\tilde{C}_{DN}^\beta c_{AN}^\beta\frac{[\text{Vol}(\mathcal{V}_N^\beta)]^2}{(2\pi)^{(\beta-1)N}}=\frac{2^N\pi^{N(1-\beta/2)}}{N!}\prod_{1\leq i\leq N}\frac{1}{[\Gamma(\beta i/2)]^2},\\
C_{RN}^\beta&=\tilde{C}_{RN}^\beta c_{SN}^\beta\int U^\dagger dU=\tilde{C}_{RN}^\beta c_{SN}^\beta\text{Vol}(\mathcal{V}_N^\beta)=\frac{2^{N[\beta(N+3)-4]/4}}{N!}\prod_{1\leq i\leq N}\frac{1}{\Gamma(\beta i/2)}.
}[EqnC]
The volume of the Stiefel manifold $\mathcal{V}_{N,N}^\beta\equiv\mathcal{V}_N^\beta$ is given in \eqref{EqnStiefel} \cite{ratnarajah2004jacobians}.  Note that the normalization constants can also be obtained from well-known Selberg-like integrals \cite{mehta2004random}.  For example, one has
\eqn{C_{DN}^\beta=2^N\prod_{1\leq i\leq N}\frac{\Gamma(\beta/2+1)}{\Gamma(\beta i/2+1)\Gamma(\beta i/2)},}
for the Dirac case.  This identity can be obtained by exploiting the properties of the function $\Gamma(x)$.

\subsection{Jacobian}\label{SsJacobian}

In terms of the dimensionless Dirac and Majorana neutrino mass matrices $\hat{M}_D$ and $\hat{M}_R$, the light neutrino mass matrix is given by
\eqn{M_\nu=\sqrt{2}\Lambda_\nu\hat{M}_\nu=-\frac{\sqrt{2}\Lambda_D^2}{\Lambda_R}\hat{M}_D\hat{M}_R^{-1}\hat{M}_D^T,}
where $\hat{M}_\nu$ is the dimensionless light neutrino mass matrix and $\Lambda_\nu=\Lambda_D^2/\Lambda_R$ is the light neutrino mass scale.  Hence $\hat{M}_\nu=-\hat{M}_D\hat{M}_R^{-1}\hat{M}_D^T$.  To determine the pdf for $\hat{M}_\nu$ from the pdfs of $\hat{M}_D$ and $\hat{M}_R$, it is necessary to compute the norm of the determinant of the Jacobian matrix corresponding to the appropriate change of variables.  Since $\hat{M}_R$ and $\hat{M}_\nu$ have the same number of independent parameters, this computation simplifies greatly when the change of variables is chosen from $(\hat{M}_D,\hat{M}_R)$ to $(\hat{M}_D,\hat{M}_\nu)$.  Thus the Jacobian matrix $J^\beta$ is given schematically by
\eqn{J^\beta=\left[\frac{\partial\hat{M}_R}{\partial\hat{M}_\nu}\right]=-\left[\frac{\partial(\hat{M}_D^T\hat{M}_\nu^{-1}\hat{M}_D)}{\partial\hat{M}_\nu}\right].}[EqnJac]

This Jacobian can be easily obtained with the help of the wedge product technique \cite{muirhead2009aspects}.  Indeed, since for a symmetric matrix $X=CYC^T$ where $C$ is a constant non-singular matrix, the measure satisfies $dX=CdYC^T$ and the wedge product leads to
\eqn{(dX)\equiv\bigwedge_{1\leq i\leq j\leq N}dX_{ij}=(CdYC^T)=p(C)(dY),}
where $p(C)$ is a polynomial in $C$.  Using $C=C_2C_1$, one obtains that $p(C_2C_1)=p(C_1)p(C_2)$.  The only function satisfying this condition is a positive power of the determinant, hence $p(C)=(\det C)^k$ with $k\geq0$.  To obtain $k$, it suffices to choose a simple matrix, for example $C=\text{diag}(c,1,\cdots,1)$ for which
\eqn{X=\left(\begin{array}{cccc}c^2y_{11}&cy_{12}&\cdots&cy_{1n}\\cy_{12}&y_{22}&\cdots&y_{2n}\\\vdots&\vdots&\ddots&\vdots\\cy_{1n}&y_{2n}&\cdots&y_{nn}\end{array}\right).}
Therefore, the wedge product for a real matrix leads to
\eqn{(dX)=(CdYC^T)\equiv\bigwedge_{1\leq i\leq j\leq N}(CdYC^T)_{ij}=c^{N+1}\bigwedge_{1\leq i\leq j\leq N}dY_{ij}=c^{N+1}(dY),}
while for a complex matrix the result becomes
\footnotesize
\eqn{(dX)=(CdYC^T)\equiv\bigwedge_{1\leq i\leq j\leq N}\text{Re}(CdYC^T)_{ij}\wedge\text{Im}(CdYC^T)_{ij}=c^{2(N+1)}\bigwedge_{1\leq i\leq j\leq N}d\text{Re}Y_{ij}\wedge d\text{Im}Y_{ij}=c^{2(N+1)}(dY).}
\normalsize
Thus one concludes that the measure satisfies
\eqn{dX=|\det C|^{\beta(N+1)}dY,}[EqnMuirhead]
\textit{i.e.} the positive power is $k=\beta(N+1)$.

It is now straightforward to obtain the Jacobian of interest.  Indeed, using \eqref{EqnMuirhead} one has
\eqn{d\hat{M}_R=-\hat{M}_D^Td\hat{M}_\nu^{-1}\hat{M}_D=\hat{M}_D^T\hat{M}_\nu^{-1}d\hat{M}_\nu\hat{M}_\nu^{-1}\hat{M}_D=|\det(\hat{M}_\nu^{-1}\hat{M}_D)|^{\beta(N+1)}d\hat{M}_\nu}
where the identity $d\hat{M}_\nu^{-1}=-\hat{M}_\nu^{-1}d\hat{M}_\nu\hat{M}_\nu^{-1}$ was used.

Finally, the full measure is simply
\eqn{d\hat{M}_Dd\hat{M}_R=\left|\frac{\det\hat{M}_D}{\det\hat{M}_\nu}\right|^{\beta(N+1)}d\hat{M}_Dd\hat{M}_\nu.}[EqnMeasure]
It is now straightforward to verify that the measure $d\hat{M}_\nu$ is invariant under every transformation $\hat{M}_\nu\to\hat{M}_\nu'=U_\nu\hat{M}_\nu U_\nu^T$ where $U_\nu\in\{O(N),U(N)\}$ for $\beta=\{1,2\}$ respectively.  Indeed, since $d\hat{M}_D'=d\hat{M}_D$ and $d\hat{M}_R'=d\hat{M}_R$ for $\hat{M}_D'=U_L\hat{M}_DU_R^\dagger$ and $\hat{M}_R'=U\hat{M}_RU^T$ respectively, one obtains
\eqna{
0&=d\hat{M}_D'd\hat{M}_R'-d\hat{M}_Dd\hat{M}_R=\left|\frac{\det\hat{M}_D'}{\det\hat{M}_\nu'}\right|^{\beta(N+1)}d\hat{M}_D'd\hat{M}_\nu'-\left|\frac{\det\hat{M}_D}{\det\hat{M}_\nu}\right|^{\beta(N+1)}d\hat{M}_Dd\hat{M}_\nu\\
&=\left|\frac{\det\hat{M}_D}{\det\hat{M}_\nu}\right|^{\beta(N+1)}d\hat{M}_D(d\hat{M}_\nu'-d\hat{M}_\nu),
}
with $\hat{M}_\nu'=-\hat{M}_D'\hat{M}_R'^{-1}\hat{M}_D'^T=U_L\hat{M}_\nu U_L^T$ and $U=U_R^*$.  Therefore, the measure $d\hat{M}_\nu'=d\hat{M}_\nu$ for every transformation in the appropriate group as stated above.  Hence $d\hat{M}_\nu$ can be written in terms of its singular values in exactly the same way as $d\hat{M}_R$ in \eqref{EqndMR}.

\subsection{Probability Density Function}\label{SsPDF}

Now that the norm of the determinant of the Jacobian matrix \eqref{EqnJac} has been found, it is simple to determine the pdf $P_\nu(\hat{M}_\nu)d\hat{M}_\nu$.  Indeed, with the help of the change of variables from $(\hat{M}_D,\hat{M}_R)$ to $(\hat{M}_D,\hat{M}_\nu)$, the pdf is calculated from $P_D(\hat{M}_D)P_R(\hat{M}_R)d\hat{M}_Dd\hat{M}_R=P_{D\nu}(\hat{M}_D,\hat{M}_\nu)d\hat{M}_Dd\hat{M}_\nu$ and the measure \eqref{EqnMeasure}, which leads to
\eqn{P_{D\nu}(\hat{M}_D,\hat{M}_\nu)d\hat{M}_Dd\hat{M}_\nu=P_D(\hat{M}_D)P_R(-\hat{M}_D^T\hat{M}_\nu^{-1}\hat{M}_D)|\det J^\beta|d\hat{M}_Dd\hat{M}_\nu.}[EqnPDFMDMnu]
The pdf $P_\nu(\hat{M}_\nu)d\hat{M}_\nu$ is obtained from \eqref{EqnPDFMDMnu} after marginalizing over the variables $\hat{M}_D$, which gives
\eqn{P_\nu(\hat{M}_\nu)d\hat{M}_\nu=\tilde{C}_{\nu N}^\beta\left[\int e^{-\text{tr}(\hat{M}_D^\dagger\hat{M}_D)-\text{tr}(\hat{M}_D^\dagger\hat{M}_\nu^{-\dagger}\hat{M}_D^*\hat{M}_D^T\hat{M}_\nu^{-1}\hat{M}_D)}\left|\frac{\det\hat{M}_D}{\det\hat{M}_\nu}\right|^{\beta(N+1)}d\hat{M}_D\right]d\hat{M}_\nu,}[EqnPDFnu]
where \eqref{EqntC} give the normalization constant as
\eqn{\tilde{C}_{\nu N}^\beta=\tilde{C}_{DN}^\beta\tilde{C}_{RN}^\beta=\frac{2^{\beta N(N-1)/4}}{\pi^{\beta N(3N+1)/4}}.}[EqntCnu]

In terms of the light neutrino mass matrix singular values $\hat{m}_{\nu,i}$ and group variables $U_\nu$, the neutrino pdf \eqref{EqnPDFnu} becomes
\small
\eqna{
P_\nu(\hat{m}_\nu)d\hat{m}_\nu\frac{U_\nu^\dagger dU_\nu}{\text{Vol}(\mathcal{V}_N^\beta)}&=\left[\int P_D(\hat{M}_D)P_R(-\hat{M}_D'^TD_\nu^{-1}\hat{M}_D')|\det J'^\beta|d\hat{M}_D\right]\\
&\phantom{=}\hspace{0.5cm}\times c_{SN}^\beta\prod_{1\leq i<j\leq N}|\hat{m}_{\nu,i}^\beta-\hat{m}_{\nu,j}^\beta|\prod_{1\leq i\leq N}|\hat{m}_{\nu,i}|^{\beta-1}d\hat{m}_{\nu,i}\frac{U_\nu^\dagger dU_\nu}{\text{Vol}(\mathcal{V}_N^\beta)}\\
&=C_{\nu N}^\beta I_N^\beta(\hat{m}_{\nu,1},\cdots,\hat{m}_{\nu,N})\prod_{1\leq i<j\leq N}|\hat{m}_{\nu,i}^\beta-\hat{m}_{\nu,j}^\beta|\prod_{1\leq i\leq N}|\hat{m}_{\nu,i}|^{-(\beta N+1)}d\hat{m}_{\nu,i}\frac{U_\nu^\dagger dU_\nu}{\text{Vol}(\mathcal{V}_N^\beta)},
}[EqnPDFnuSVG]
\normalsize
where the normalization constant is obtained from \eqref{EqntCnu} and is given by
\eqn{C_{\nu N}^\beta=\frac{\tilde{C}_{\nu N}^\beta c_{SN}^\beta c_{AN}^\beta}{2^N}\frac{[\text{Vol}(\mathcal{V}_N^\beta)]^3}{(2\pi)^{(\beta-1)N}}=\frac{2^{N[\beta(N+3)-4]/4}}{N!}\prod_{1\leq i\leq N}\frac{\Gamma(\beta/2+1)}{\Gamma(\beta i/2+1)[\Gamma(\beta i/2)]^2},}[EqnCnu]
while the remaining function is
\small
\eqna{
I_N^\beta(t_1,\cdots,t_N)&=\frac{2^N}{\text{Vol}(\mathcal{V}_N^\beta)}\int_{U_L,U_R\in\mathcal{V}_N^\beta}\int_0^\infty e^{-\sum_{1\leq i,j\leq N}|\sum_{1\leq k\leq N}t_k^{-1}(U_L)_{ki}(U_L)_{kj}|^2\hat{m}_{D,i}^2\hat{m}_{D,j}^2}\\
&\phantom{=}\hspace{0.5cm}\times\prod_{1\leq i<j\leq N}|\hat{m}_{D,i}^2-\hat{m}_{D,j}^2|^\beta\prod_{1\leq i\leq N}e^{-\hat{m}_{D,i}^2}|\hat{m}_{D,i}|^{\beta(N+2)-1}d\hat{m}_{D,i}\frac{(U_L^\dagger dU_L)'(U_R^\dagger dU_R)'}{\text{Vol}(\mathcal{V}_N^\beta)/(2\pi)^{(\beta-1)N}}\\
&=\int_{U\in\mathcal{V}_N^\beta}\int_0^\infty\prod_{1\leq i<j\leq N}|x_i-x_j|^\beta e^{-2|\sum_{1\leq k\leq N}t_k^{-1}U_{ki}U_{kj}|^2x_ix_j}\\
&\phantom{=}\hspace{0.5cm}\times\prod_{1\leq i\leq N}x_i^{\beta(N+2)/2-1}e^{-x_i(1+|\sum_{1\leq j\leq N}t_j^{-1}U_{ji}^2|^2x_i)}dx_i\frac{(U^\dagger dU)'}{\text{Vol}(\mathcal{V}_N^\beta)/(2\pi)^{(\beta-1)N}}.
}[EqnI]
\normalsize
In deriving \eqref{EqnPDFnuSVG}, the change of variables $\hat{M}_\nu=U_\nu D_\nu U_\nu^T$ where $D_\nu=\text{diag}(\hat{m}_{\nu,1},\cdots,\hat{m}_{\nu,N})$ was used.  The invariance of the measure $d\hat{M}_D$ under the appropriate transformations was also necessary.  To rewrite the function $I_N^\beta$, the Dirac neutrino mass matrix elements was expressed in terms of its singular values and its group variables.  The group variables for $U_R$ were integrated over straightforwardly while the group variables for $U_L$ (minus the phases, hence the prime) remain due to their complicated coupling with the singular values.  Finally, the change of variables $x_i=\hat{m}_{D,i}^2$ and $U_L=U$ were done to simplify the integral $I_N^\beta$.

Note that integration over the light neutrino group variables $U_\nu$ in \eqref{EqnPDFnuSVG} is straightforward, as in \eqref{EqnPDFdmh}, leading to
\eqn{P_\nu(\hat{m}_\nu)d\hat{m}_\nu=C_{\nu N}^\beta I_N^\beta(\hat{m}_{\nu,1},\cdots,\hat{m}_{\nu,N})\prod_{1\leq i<j\leq N}|\hat{m}_{\nu,i}^\beta-\hat{m}_{\nu,j}^\beta|\prod_{1\leq i\leq N}|\hat{m}_{\nu,i}|^{-(\beta N+1)}d\hat{m}_{\nu,i},}[EqnPDFnuSV]
for the singular value pdf, with the normalization constant \eqref{EqnCnu}.  From \eqref{EqnPDFnuSVG}, the pdf for the group variables $U_\nu$ is thus uniquely determined by the appropriate group (Haar) measure (minus the phases).  This behavior is problematic since there must be a correlation between the neutrino masses (\textit{i.e.} the singular values of $\hat{M}_\nu$) and the neutrino mass eigenstates (\textit{i.e.} the singular vectors of $\hat{M}_\nu$ which are given by the columns of the neutrino mixing matrix $U_\nu$).  For example, the mostly-electronic neutrino must be the lightest (normal hierarchy) or the second lightest (inverted hierarchy).  This simple observation has important consequences in the analysis of the physical case appropriate for the SM.

For general $N$, the function \eqref{EqnI} makes the analysis of the see-saw ensemble pdf \eqref{EqnPDFnuSV} quite intricate since it is not only a random matrix theory pdf, it is a random matrix theory pdf that cannot be written as an analytic function (apart for the $N=1$ case, to the best of our knowledge).

For real matrices, none of the group variables in \eqref{EqnI} can be straightforwardly integrated.  For complex matrices however, the integration over group variables does not include some of the phases, hence the prime as explained in appendix \ref{SMeasures}.  This can also be seen by using an adequate parametrization of the matrix $U$ in \eqref{EqnI}.  Indeed, following in part \cite{1751-8121-43-38-385306,:/content/aip/journal/jmp/53/1/10.1063/1.3672064}, it is possible to parametrize any $N\times N$ unitary matrix as
\eqn{U=\prod_{1\leq j<k\leq N}\exp(i\phi_{jk}P_k)\exp(i\theta_{jk}\Sigma_{jk})\prod_{1\leq j\leq N}\exp(i\varphi_jP_j),}
where the matrices $P_j$ and $\Sigma_{jk}$ are given by
\eqn{P_{j,ik}=\delta_{ji}\delta_{jk},\quad\quad\Sigma_{jk,i\ell}=-i\delta_{ji}\delta_{k\ell}+i\delta_{j\ell}\delta_{ki},}
and the range of the $N^2$ mixing angles $\theta_{jk}$ and phases $\phi_{jk}$ and $\varphi_j$ are
\eqn{\theta_{jk}\in[0,\pi/2),\quad\quad\phi_{jk}\in[0,2\pi),\quad\quad\varphi_j\in[0,2\pi).}
With this parametrization it is clear that the integrand in \eqref{EqnI} is independent of the $N$ phases $\varphi_j$, explaining why they are not integrated over in \eqref{EqnI}.


\section{Properties of the See-Saw Ensemble}\label{SProperties}

This section discusses the general properties of the see-saw ensemble pdf \eqref{EqnPDFnuSV}.  The asymptotic behaviors at $\hat{m}_{\nu,i}\to0$ and $\hat{m}_{\nu,i}\to\infty$ are investigated.  Moreover, the $N=1$ case as well as the large $N$ case are studied analytically.

\subsection{Asymptotic Behaviors}\label{SsAsymptotics}

First, being a pdf, \eqref{EqnPDFnuSV} implies the following identity for $\beta=1$ or $\beta=2$,
\eqn{\int_{\hat{m}_{\nu,\text{min}}}^\infty I_N^\beta(\hat{m}_{\nu,1},\cdots,\hat{m}_{\nu,N})\prod_{1\leq i<j\leq N}|\hat{m}_{\nu,i}^\beta-\hat{m}_{\nu,j}^\beta|\prod_{1\leq i\leq N}|\hat{m}_{\nu,i}|^{-(\beta N+1)}d\hat{m}_{\nu,i}=\frac{1}{C_{\nu N}^\beta},}
where $\hat{m}_{\nu,\text{min}}=-\infty$ for $\beta=1$ (eigenvalues) and $\hat{m}_{\nu,\text{min}}=0$ for $\beta=2$ (singular values).

The pdf \eqref{EqnPDFnuSV} is difficult to study analytically for general $N$ since the function \eqref{EqnI} is hard to evaluate generally.  It is nevertheless possible to investigate the asymptotic behaviors of the pdf \eqref{EqnPDFnuSV} for general $N$ as follows.

First, as $\hat{m}_{\nu,i}\to\pm\infty$ for a fixed $i$, one has
\eqn{P_\nu(\hat{m}_\nu)=k_\infty|\hat{m}_{\nu,i}|^{\beta (N-1)-(\beta N+1)}\left[1+\mathcal{O}(|\hat{m}_{\nu,i}|^{-1})\right]=k_\infty|\hat{m}_{\nu,i}|^{-(\beta+1)}\left[1+\mathcal{O}(|\hat{m}_{\nu,i}|^{-1})\right].}
Therefore, as long as the asymptotic expansion converges uniformly, the average singular value (and all moments greater than $1$) is not well-defined for $\beta=1$ while the standard deviation (and all moments greater than $2$) is not well-defined for $\beta=2$.  Here the constant $k_\infty$ depends on the other singular values and on $N$ and is thus difficult to calculate analytically.

The limiting behavior at $\hat{m}_{\nu,i}\to0$ is obtained from the function $I_N^\beta$.  From \eqref{EqnI} and the rescaling $x_j=|\hat{m}_{\nu,i}|y_j$, one gets
\eqna{
I_N^\beta(\hat{m}_{\nu,1},\cdots,\hat{m}_{\nu,N})&\sim|\hat{m}_{\nu,i}|^{\beta N(2N+1)/2}\int_{U\in\mathcal{V}_N^\beta}\int_0^\infty\prod_{1\leq j<k\leq N}|y_j-y_k|^\beta e^{-2\hat{m}_{\nu,i}^2|\sum_{1\leq\ell\leq N}\hat{m}_{\nu,\ell}^{-1}U_{\ell j}U_{\ell k}|^2y_jy_k}\\
&\phantom{=}\hspace{0.5cm}\times\prod_{1\leq j\leq N}y_j^{\beta(N+2)/2-1}e^{-|\hat{m}_{\nu,i}|y_j(1+|\hat{m}_{\nu,i}||\sum_{1\leq k\leq N}\hat{m}_{\nu,k}^{-1}U_{kj}^2|^2y_j)}dy_j(U^\dagger dU)',
}
and that implies that for $\hat{m}_{\nu,i}\to0$, the pdf behaves as
\eqn{P_\nu(\hat{m}_\nu)=k_0|\hat{m}_{\nu,i}|^{\beta N(2N-1)/2-1}\left[1+\mathcal{O}(|\hat{m}_{\nu,i}|)\right].}
Therefore the pdf vanishes at $\hat{m}_{\nu,i}=0$ unless $N=1$.  Again the constant $k_0$ is a function of the remaining singular values and $N$ that is hard to evaluate exactly.

With the help of both asymptotic behaviors, it is straightforward to conclude that the most probable singular value is a finite number different than zero, unless $N=1$.  In other words, for $N\geq2$ the see-saw ensemble prefers the lightest neutrino to be massive.

\subsection{The $N=1$ case}\label{SsN1}

\begin{figure}[!t]
\centering
\resizebox{15cm}{!}{
\includegraphics{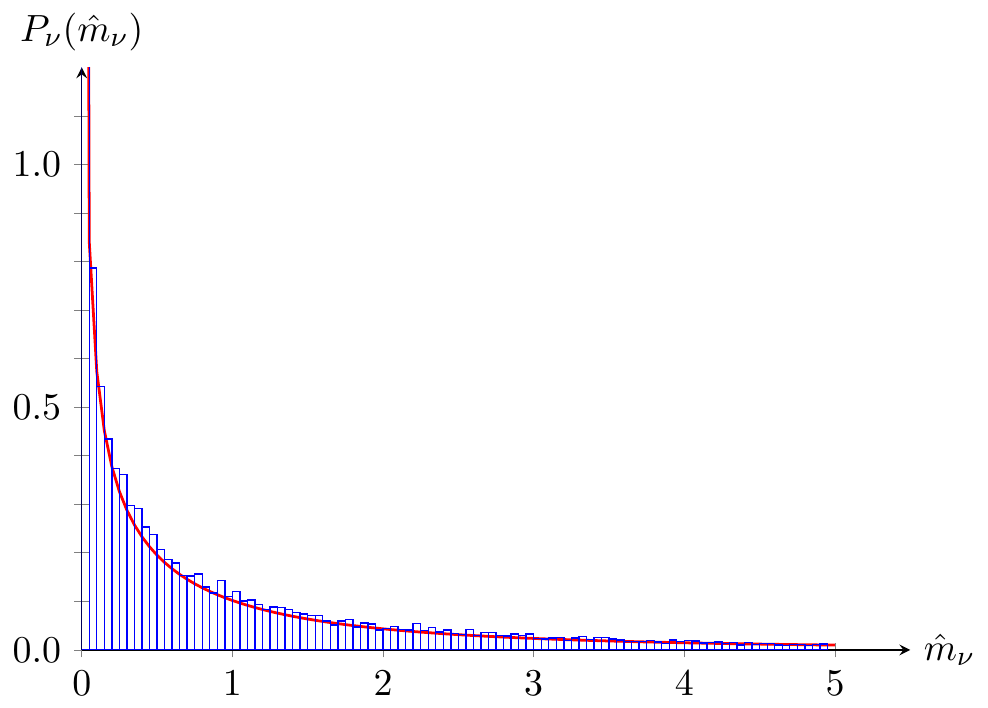}
\hspace{2cm}
\includegraphics{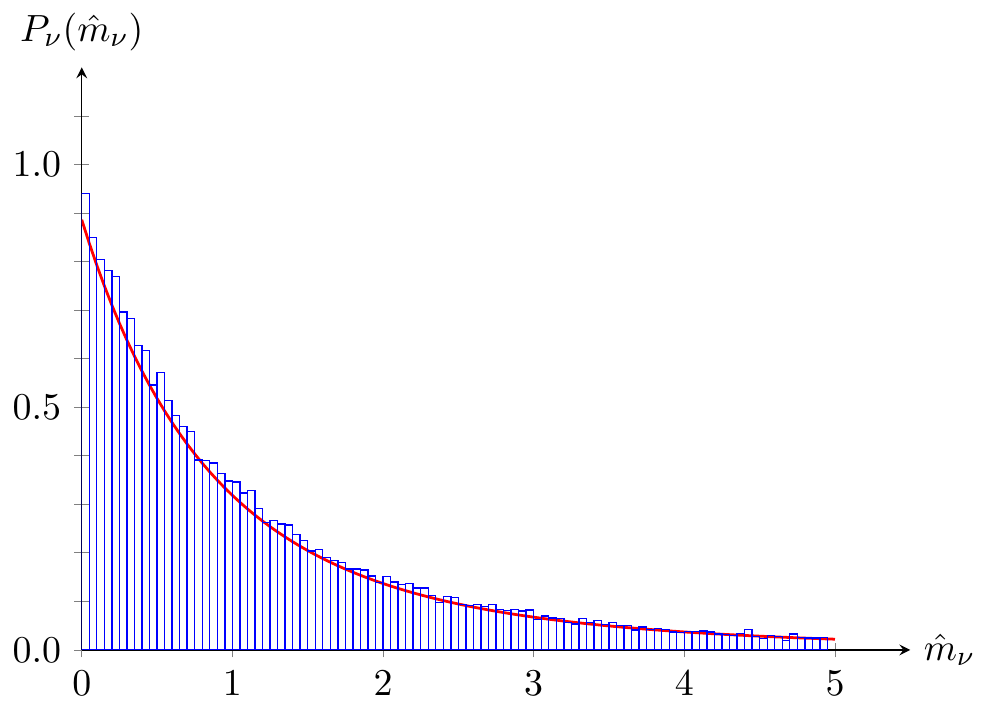}
}
\caption{Probability density function for the see-saw ensemble with $N=1$.  The red curve corresponds to the analytic result while the histogram corresponds to numerical results (with $5\times10^4$ dimensionless light neutrino mass matrices generated).  The left panel shows matrices with real elements ($\beta=1$) while the right panel shows matrices with complex elements ($\beta=2$).}
\label{FigPDFnuSVN1}
\end{figure}

For now, the only analytic case with finite $N$ corresponds to $N=1$, for which the joint pdf is given by
\eqn{P_\nu(\hat{m}_\nu)d\hat{m}_\nu=\frac{\pi^{1-\beta/2}}{2^{\beta/2+1}}\frac{\Gamma(3\beta/2)}{[\Gamma(\beta/2)]^3}|\hat{m}_\nu|^{(\beta-2)/2}U(3\beta/4,1/2,\hat{m}_\nu^2/4),}[EqnPDFnuSVN1]
where $U(a,b,z)$ is the confluent hypergeometric function of the second kind.  The joint pdf \eqref{EqnPDFnuSVN1} can be obtained analytically because the function \eqref{EqnI} is simple in the $N=1$ case.  The plot of \eqref{EqnPDFnuSVN1} is shown in figure \ref{FigPDFnuSVN1} along with numerical results.  The agreement between the two approaches is clear.

Figure \ref{FigPDFnuSVN1} shows that the most probable value for the singular value is $\hat{m}_\nu=0$.  This is easily verified by looking at $\partial P_\nu(\hat{m}_\nu)/\partial\hat{m}_\nu$ which is always negative (positive) for positive (negative) $\hat{m}_\nu$.  Moreover, for $\beta=2$, the pdf of the light neutrino phase is the Haar measure which is flat and therefore uninteresting.  Also, from the limiting behaviors, which can be obtained directly from
\eqn{
U(3\beta/4,1/2,z)=\left\{\begin{array}{lll}z^{-3\beta/4}\left[1+\mathcal{O}(z^{-1})\right]&\text{for}&z\to\infty\\\frac{\pi^{1/2}}{\Gamma[(3\beta+2)/4]}\left[1+\mathcal{O}(z^{1/2})\right]&\text{for}&z\to0\end{array}\right.,
}
none of the moments exist except for the average value of the singular value for $\beta=2$.  A simple computation using the following Mellin transforms,
\eqn{\int_0^\infty z^{\lambda-1}U(a,b,z)dz=\frac{\Gamma(\lambda)\Gamma(a-\lambda)\Gamma(\lambda-b+1)}{\Gamma(a)\Gamma(a-b+1)}\quad\quad\text{for}\quad\quad\text{max}(b-1,0)<\lambda<a,}
shows it is given by $\langle\hat{m}_\nu\rangle_{N=1}^{\beta=2}=\sqrt{\pi}$.  Since \eqref{EqnPDFnuSVN1} is an even function of $\hat{m}_\nu$, one can nevertheless formally define the average value for $\beta=1$, and it is $\langle\hat{m}_\nu\rangle_{N=1}^{\beta=1}=0$.  These observations agree with the general analysis of section \ref{SsAsymptotics}.

\subsection{The large $N$ case}\label{SsLargeN}

For usual Gaussian ensembles, both the $N=1$ pdfs and the level densities at large $N$ are simple.  Indeed, for the usual Gaussian random matrix theory, one can show the celebrated Wigner's semicircle law using the moment method or the resolvent method.  It is therefore plausible that the level density at large $N$ for the see-saw ensemble is also a simple analytic function.  Moreover, from the $1/N$ expansion, the large $N$ case could shed some light on the physical case of $N=3$ neutrinos, as occurs for example with quantum chromodynamics.  It is however impossible here to use the moment method or the resolvent method since both methods rely on the computation of the moments and most moments do not exist for the see-saw ensemble, as argued above.

Another option is to translate the problem to a Coulomb-like gas and take the thermodynamic limit, which effectively corresponds to the limit $N\to\infty$.  To do so, the singular value pdf \eqref{EqnPDFnuSV} is re-expressed as a thermal system $e^{-H_{\nu N}^\beta}$ where the temperature is fixed.  In this picture, the Hamiltonian for the see-saw ensemble is obtained from \eqref{EqnPDFnuSV} and is given by
\eqn{H_{\nu N}^\beta=-\sum_{1\leq i<j\leq N}\ln|x_i^\beta-x_j^\beta|+(\beta N+1)\sum_{1\leq i\leq N}\ln|x_i|-\ln I_N^\beta(x_1,\cdots,x_N),}[EqnHN]
where the positions of the $N$ charged particles are given by $x_i$.  It is important to note that the Hamiltonian \eqref{EqnHN} corresponds to a Coulomb gas only for $\beta=1$.  Indeed, the characteristic logarithmic two-dimensional Coulomb potential occurs only for $\beta=1$.

The thermodynamic limit is computable with the help of the level density of singular values $\rho_{\nu N}^\beta(x)$, which is defined as the following correlation function,
\eqn{\rho_{\nu N}^\beta(x)=N\int P_\nu(x,\hat{m}_{\nu,2},\cdots,\hat{m}_{\nu,N})\prod_{2\leq i\leq N}d\hat{m}_{\nu,i},}
such that $\int\rho_{\nu N}^\beta(x)dx=N$ the number of charged particles.  Assuming the distance between charged particles decreases as $N$ tends to infinity, the Hamiltonian \eqref{EqnHN} in the large $N$ limit can be approximated by a continuum of singular values with Hamiltonian,
\eqna{
H_{\nu N}^\beta&=-\frac{1}{2}\int dxdy\,\rho_{\nu N}^\beta(x)\rho_{\nu N}^\beta(y)\ln|x^\beta-y^\beta|+\int dx\,\rho_{\nu N}^\beta(x)V_{\nu N}^\beta,\\
V_{\nu N}^\beta&=(\beta N+1)\ln|x|+\cdots,}[EqnHinfty]
where the ellipses represents the contribution from the integral $I_N^\beta$ to the potential $V_{\nu N}^\beta$.  Extremizing the Hamiltonian \eqref{EqnHinfty} with respect to the level density of singular values subject to the constraint $\int\rho_{\nu N}^\beta(x)dx=N$ leads to an equation for the level density of singular values which can usually be solved analytically.  This is done in appendix \ref{SCircle} for real and complex arbitrary matrices as well as real and complex symmetric matrices with Gaussian ensembles.  This is however not the case here due to the complexity of the integral $I_N^\beta$.  Indeed, for usual Gaussian ensembles, the potential $V_N^\beta\propto x^2$ depends only on one variable and extremization of the Hamiltonian is straightforward (see appendix \ref{SCircle}).  For the see-saw ensemble however, the potential $V_{\nu N}^\beta$ is a complicated integral function of several variables, thus extremizing the Hamiltonian is not so simple.  Moreover, since $e^{-1/x^2}$ does not have an expansion around $x=0$, it is not possible to compute the first contributions to the potential from the integral $I_N^\beta$.

To proceed, it is proposed to approximate the problem by computing the level densities in the thermodynamic limit for both the Dirac and Majorana neutrino mass matrices first, and then marginalizing to get the level density of singular values for the see-saw ensemble.  Since interchanging the order of the steps gets rid of all complications related to the group variables, this technique can at best give an approximation to the true level density of singular values.  Although a comparison between the analytical approximation obtained with this technique and numerical results shows that the approximation is quite good, it will be argued that the results are wrong.

\begin{figure}[!t]
\centering
\resizebox{15cm}{!}{
\includegraphics{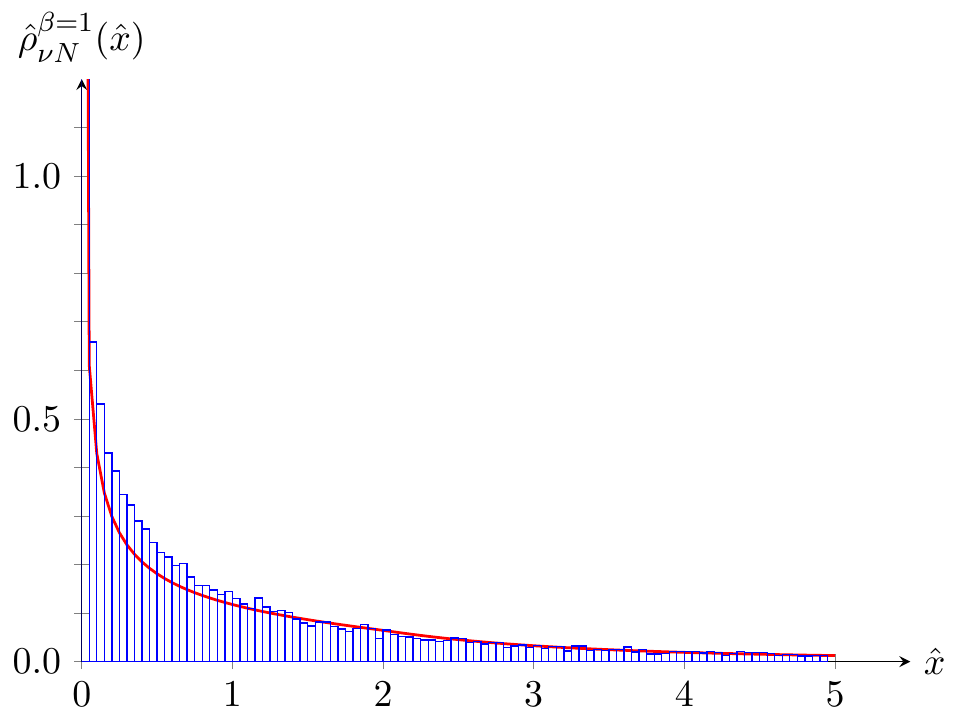}
\hspace{2cm}
\includegraphics{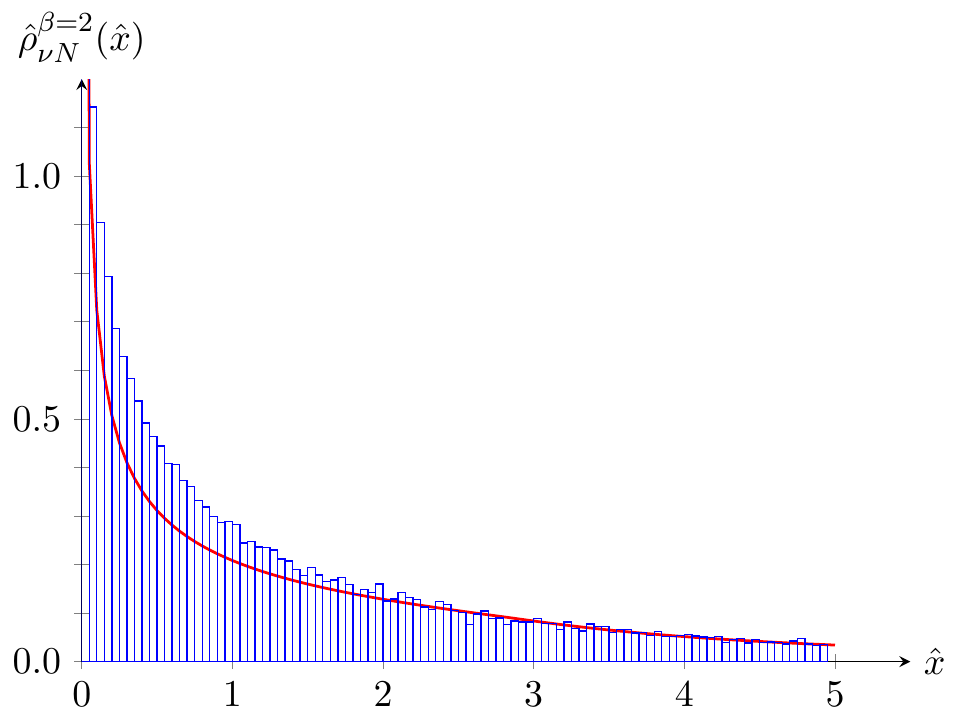}
}
\caption{Level density at large $N$ for the see-saw ensemble.  The red curve corresponds to the approximated analytic result while the histogram corresponds to numerical results (with $10^3$ singular values generated from $50\times50$ matrices).  The left panel shows matrices with real elements ($\beta=1$) while the right panel shows matrices with complex elements ($\beta=2$).  Although the fits seem good, the approximated analytic result for $\beta=2$ is without a doubt not correct.}
\label{Figrho}
\end{figure}

The level densities for the Dirac (arbitrary) and Majorana (symmetric) matrices are given in \eqref{EqnrhoSAsoln}.  Following the same procedure as in section \ref{SsPDF} and marginalizing with the approximation $\hat{x}_R=\hat{x}_D^2/\hat{x}$ where $\hat{x}\equiv\hat{x}_\nu$, which is the analog of $\hat{M}_R=-\hat{M}_D^T\hat{M}_\nu^{-1}\hat{M}_D$, gives
\eqn{\hat{\rho}_{\nu N}^\beta(\hat{x})=\int d\hat{x}_D\,\hat{\rho}_D(\hat{x}_D)\hat{\rho}_R(\hat{x}_D^2/\hat{x})|\det\hat{J}|.}
Here the normalized quantities $\hat{x}=x/\sqrt{N}$ and $\hat{\rho}(\hat{x})=\rho(x)/\sqrt{N}$ are introduced in appendix \ref{SCircle}.  Moreover, the change of variables leads to the Jacobian $|\det\hat{J}|=\hat{x}_D^2/\hat{x}^2$.  The solution to the marginalization procedure of the two level densities is given by
\eqn{\hat{\rho}_{\nu N}^\beta(\hat{x})=\left\{\begin{array}{lll}\frac{\beta^{3/4}}{10\pi^{5/2}|\hat{x}|^{1/2}}\left[\Gamma\left(-\frac{1}{4}\right)^2{}_3F_2\left(-\frac{1}{4},\frac{1}{4},\frac{3}{4};\frac{1}{2},\frac{9}{4};\frac{\hat{x}^2}{4\beta}\right)\right.&&\\\hspace{1cm}\left.-\frac{10}{21}\frac{|\hat{x}|}{2\beta^{1/2}}\Gamma\left(\frac{1}{4}\right)^2{}_3F_2\left(\frac{1}{4},\frac{3}{4},\frac{5}{4};\frac{3}{2},\frac{11}{4};\frac{\hat{x}^2}{4\beta}\right)\right]&\text{for}&|\hat{x}|<2\sqrt{\beta}\\\frac{\beta^{3/2}}{\pi\hat{x}^2}{}_3F_2\left(-\frac{1}{2},\frac{3}{4},\frac{5}{4};\frac{3}{2},2;\frac{4\beta}{\hat{x}^2}\right)&\text{for}&|\hat{x}|\geq2\sqrt{\beta}\end{array}\right..}[Eqnrho]
The continuity at $\hat{x}=2\sqrt{\beta}$ and the normalization $\int d\hat{x}\,\hat{\rho}_{\nu N}^\beta(\hat{x})=1$ of the level density \eqref{Eqnrho} can be verified exactly by using generalizations of the Gauss's identity ${}_2F_1(a,b;c;1)=\frac{\Gamma(c)\Gamma(c-a-b)}{\Gamma(c-a)\Gamma(c-b)}$ for $c-a-b>0$.  The level density for the see-saw ensemble \eqref{Eqnrho} is shown in figure \ref{Figrho} with a comparison to numerical results.  The behaviors of the level density around zero and infinity (more precisely $|\hat{x}|\gtrsim2$) match the numerical results quite well while the intermediate regime is not as good.  The agreement is nonetheless satisfying considering the approximations made.

The limiting behaviors at $\hat{x}\to0$ and $\hat{x}\to\pm\infty$ are
\eqn{\hat{\rho}_{\nu N}^\beta(\hat{x})=\left\{\begin{array}{lll}\frac{\beta^{3/4}\Gamma\left(-\frac{1}{4}\right)^2}{10\pi^{5/2}|\hat{x}|^{1/2}}&\text{for}&|\hat{x}|\to0\\\frac{\beta^{3/2}}{\pi\hat{x}^2}&\text{for}&|\hat{x}|\to\infty\end{array}\right.,}
and they imply that none of the moments exist although the level density is integrable.  This is consistent with the results obtained in section \ref{SsAsymptotics} from the full pdf for $\beta=1$ but not for $\beta=2$.  It is therefore expected that the match between the approximated level density \eqref{Eqnrho} and numerical results is better for $\beta=1$ compared to $\beta=2$.  Although both the real and complex cases seem to agree with the numerical results as seen in figure \ref{Figrho}, it is clear from the difference in the moments that the level density at large $N$ \eqref{Eqnrho} is wrong for $\beta=2$.  The error originates from all the simplifications made, most likely from the Jacobian of the transformation.  Indeed, the expected $\beta$-dependence of the Jacobian, as in \eqref{EqnMeasure}, does not occur due to the approximations, making the analogy with the full pdf possibly correct only for $\beta=1$.  Hence computing the level densities at large $N$ for the Dirac and Majorana neutrino mass matrices and then marginalizing to obtain the level density at large $N$ for the see-saw ensemble do not commute for $\beta=2$.

Furthermore, contrary to usual level densities, the level density for the see-saw ensemble does not have compact support due to the inverse Majorana neutrino mass matrix that appears in the marginalization.  Hence there are arbitrarily large singular values, although the level density and the mean level spacing, given by $1/\hat{\rho}_{\nu N}^\beta(\hat{x})$, show that large singular values are rare and far apart from each other.

Finally, apart from the fact that the approximated analytic level density seems wrong, it seems ill-advised to work back and compute the potential for the see-saw ensemble from the level density since it is already known that $V_{\nu N}^\beta$ depends on more than one variable.

\subsection{Comparison between $N=1$ and large $N$}

Before concluding, it is interesting to compare the results for $N=1$ and large $N$.  Indeed, although those regimes are as far apart as possible, by definition the pdf for $N=1$ is the $N=1$ (normalized) level density.  Hence, \eqref{EqnPDFnuSVN1} can be compared directly with \eqref{Eqnrho}, allowing a qualitative understanding of the dependence on $N$.  Moreover, this comparison will help show that the analysis of the previous section is most likely wrong even for $\beta=1$.

In the usual Gaussian ensemble, the comparison is between a Gaussian distribution for the pdf and Wigner's semicircle distribution for the level density at large $N$.  It is clear that the agreement between the two functions is poor.  Nevertheless, comparing qualitatively the distributions shows quite strikingly that the support of the normalized level density changes from being non-compact at $N=1$ to being compact at large $N$.

For the see-saw ensemble, it is convenient to discuss the real and complex cases separately.
\begin{figure}[!t]
\centering
\resizebox{15cm}{!}{
\includegraphics{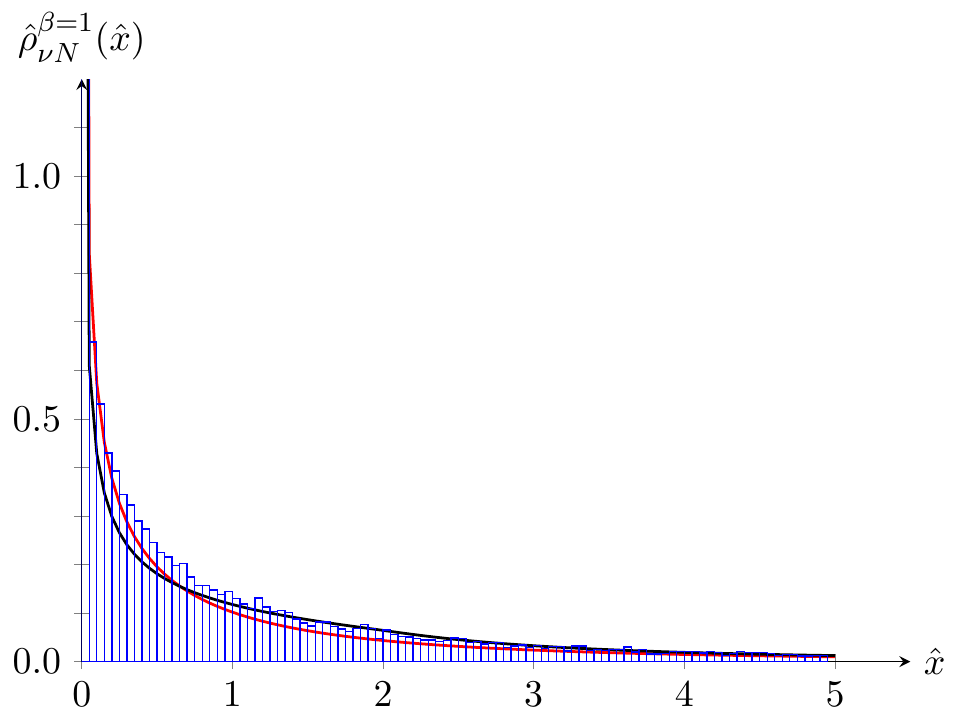}
\hspace{2cm}
\includegraphics{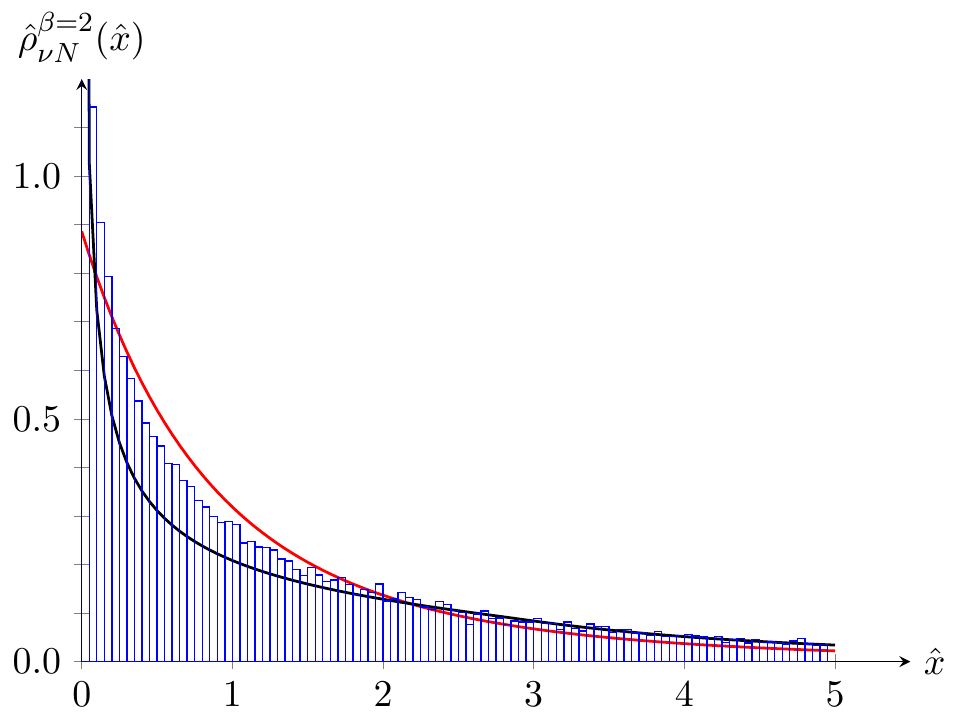}
}
\caption{Comparison between the probability density function at $N=1$ and the level density at large $N$ for the see-saw ensemble.  The red curve corresponds to probability density function, the black curve corresponds to the approximated analytic result while the histogram corresponds to the same numerical results as in figure \ref{Figrho}.  The left panel shows matrices with real elements ($\beta=1$) while the right panel shows matrices with complex elements ($\beta=2$).}
\label{FigN1Nlarge}
\end{figure}

For the complex case ($\beta=2$), it is already known that the approximated analytic level density at large $N$ is not correct, hence it is only possible to compare the analytic pdf \eqref{EqnPDFnuSVN1} with the numerical results in figure \ref{Figrho} at large $N$.  The comparison as shown in the right panel of figure \ref{FigN1Nlarge} demonstrates that the pdf at $N=1$ agrees well with the level density at large $N$ for $\hat{x}\gtrsim2$ only.  This shows qualitatively that the first moment of the level density at large $N$ exists, as expected.  Moreover, it is clear that as $N$ increases, the number of normalized singular values around zero changes from a constant for $N=1$ to possibly become infinite at large $N$.  Finally, as already mentioned, the support stays non-compact even as $N\to\infty$.

The real case ($\beta=1$) is more interesting.  Since the approximated analytic level density at large $N$ is consistent with the moments and the numerical results, it is possible to compare the analytic results of \eqref{EqnPDFnuSVN1} and \eqref{Eqnrho} as well as the numerical results of figure \ref{Figrho}, as shown in the left panel of figure \ref{FigN1Nlarge}.  A quick glance at figure \ref{FigN1Nlarge} is enough to realize that although the pdf at $N=1$ and the approximated analytic result for the level density at large $N$ are quite alike, the former is an even better fit to the numerical results for the level density at large $N$.  Hence the normalized level density at large $N$ is well approximated by the pdf at $N=1$ \eqref{EqnPDFnuSVN1}!  This observation implies that the normalized level density does not change much as $N$ increases.  It would be interesting to investigate this behavior in more detail.

In summary, the action of first computing the pdf for the see-saw ensemble from the Dirac and Majorana neutrino mass matrices and then obtaining the see-saw ensemble level density at large $N$ is not the same than first obtaining the level densities at large $N$ for the Dirac and Majorana neutrino mass matrices and then computing the see-saw ensemble level density at $N\to\infty$.  The simplifications made by interchanging the order of the steps are too extreme to generate a useful result.  From the numerical analysis, one can nevertheless conclude that $\hat{\rho}_{\nu N}^{\beta=1}$ is well approximated by $P_\nu$ at $N=1$ while $\hat{\rho}_{\nu N}^{\beta=2}(\hat{x})$ most likely behaves as $\hat{x}^{-1/2}$ when $\hat{x}\to0$ and $\hat{x}^{-3}$ when $\hat{x}\to\infty$.\footnote{The behavior of the pdf around zero suggests that the level density vanishes at zero.  If the level density peaks very close to zero before reaching zero as $\hat{x}$ decreases, this feature would be lost in the numerical analysis due to the finite binning.}


\section{Discussion and Conclusion}\label{SConclusion}

This paper investigated the see-saw ensemble originating from the anarchy principle with the help of random matrix theory.  The starting assumption was an extended SM with the type I see-saw mechanism, although the same analysis can be done for the type III see-saw mechanism since the light neutrino mass matrix is analogous.\footnote{Type II see-saw mechanism is trivial since the light neutrino mass matrix would be generated by a Gaussian ensemble.}  It is shown that the see-saw ensemble joint pdf for the singular values is a complicated function of the singular values and that it decouples from the see-saw ensemble joint pdf for the light neutrino mixing angles and phases which is simply given by the appropriate group (Haar) measure.  The fact that the light neutrino mass pdf and the light neutrino mixing matrix pdf are statistically independent is important when comparing with actual neutrino physics since it restricts the predictive power of the anarchy principle.  Its implications are briefly discussed below.

The asymptotic behavior of the pdf at large mass values suggests that, apart from the average value for complex matrices, none of the positive moments exist.  The $N=1$ case, for which the complicated function simplifies, is studied analytically.  The level density at large $N$ is then approximated but it is shown afterwards that the technique used to reach a simplified analytic result is most likely wrong.  Moreover, the numerical behavior of the level density at large $N$ found here does not seem to match with the answer found in \cite{Bai:2012zn} where an ansatz was used.

Although the physical case of complex $3\times3$ matrices will be discussed extensively elsewhere \cite{Fortin:2017iiw}, the analysis presented here is sufficient to lay out some physical implications of the see-saw ensemble:
\begin{enumerate}
\item The see-saw ensemble prefers three massive light neutrinos, a massless neutrino is forbidden;
\item In the see-saw ensemble, there does not exist a correlation between the light neutrino mass eigenstates and the mixing angles and phases, which results in a lost of predictive power when comparing with observations;
\item This lack of correlation implies that the distribution of light neutrino mixing angles and phases is simply the appropriate Haar measure, as pointed out in \cite{Haba:2000be}, and hence near-maximal mixings observed experimentally are highly probable although the connection with the mass eigenstates is lost;
\item The absence of a correlation also implies that the spectrum of light neutrino masses can exhibit the normal hierarchy, the inverted hierarchy or something else (contrary to the claim of \cite{Bai:2012zn}), as the mostly-electronic neutrino could be the heaviest one, in conflict with observations.
\end{enumerate}
The last two observations can be answered quantitatively with a thorough investigation of the see-saw ensemble pdf for $N=3$ and $\beta=2$, an analysis that will appear in subsequent work \cite{Fortin:2017iiw}.


\ack{
The authors would like to thank Patrick Desrosiers for enlightening discussions on the mathematics of random matrix theory.  This work is supported by NSERC.
}


\setcounter{section}{0}
\renewcommand{\thesection}{\Alph{section}}

\section{Measures}\label{SMeasures}

This appendix re-derives important results on the singular value decompositions of the measures encountered in the computation of the see-saw pdf.  Note that these measures were already discussed in \cite{Haba:2000be}, yet their derivation relied in part on some heuristic arguments that led to conjectures for matrices of arbitrary dimensions.  This appendix presents complete proofs making use of the wedge product technique.

\subsection{Stiefel Manifold}

In constructing the measures from the singular value decomposition theorem, a recurrent quantity of interest is the volume of Stiefel manifolds \cite{ratnarajah2004jacobians}.  The volume of the Stiefel manifold $\mathcal{V}_N^\beta\equiv\mathcal{V}_{N,N}^\beta$ is a well-known result and is simply given by
\eqn{\text{Vol}(\mathcal{V}_N^\beta)=\int_{U\in\mathcal{V}_N^\beta}U^\dagger dU=\frac{2^N\pi^{\beta N(N+1)/4}}{\prod_{1\leq i\leq N}\Gamma(\beta i/2)}.}[EqnStiefel]
This quantity helps define the group measures that appear in the singular value decomposition.

\subsection{Singular Value Decomposition}

With the knowledge of Stiefel volumes, it is now possible to obtain the measures \eqref{EqndMD} and \eqref{EqndMR} using the power of the wedge product approach.  Because of its anticommutativity, the wedge product approach allows to obtain Jacobians simply.  However, since each singular value decomposition is different \cite{RATNARAJAH2005399}, in the following each of the four cases  (real symmetric, complex symmetric, real arbitrary and complex arbitrary) are proved independently.  Note that overall minus signs are not important in the wedge product approach since only the absolute value of the Jacobian is of interest.

\subsubsection{Real and Complex Symmetric Matrix}

For a real (complex) $N\times N$ symmetric (denoted by the subscript $S$) matrix $M$, the eigen (singular) value decomposition theorem states that there exists a real (complex) matrix $U$ satisfying $U^\dagger U=\mathds{1}$ such that $M=UDU^T$ where $D$ is real diagonal, \textit{i.e.} $D=\text{diag}(\lambda_1,\cdots,\lambda_N)$, with $\lambda_N>\lambda_{N-1}>\cdots>\lambda_1$ and $\lambda_1>0$ ($\lambda_1>-\infty$).  Thus, one has
\eqn{dM=d(UDU^T)=dUDU^T+UdDU^T+UDdU^T.}[EqnSd]
Multiplying \eqref{EqnSd} by $U^\dagger$ on the left and by $U^*$ on the right leads to
\eqna{
U^\dagger dMU^*&=(U^\dagger dU)D+dD+D(U^\dagger dU)^T\\
&=(U^\dagger dU)D+dD-D(U^\dagger dU)^*\\
&=\left\{\begin{array}{lll}d\lambda_i+\lambda_i[(U^\dagger dU)_{ii}-(U^\dagger dU)_{ii}^*]&\text{for}&i=j\\\lambda_j(U^\dagger dU)_{ij}-\lambda_i(U^\dagger dU)_{ij}^*&\text{for}&i\neq j\end{array}\right.,
}[EqnSd1]
where the second and last equalities come from $U^\dagger dU=d(U^\dagger U)-dU^\dagger U=-(U^\dagger dU)^\dagger$.  From \eqref{EqnMuirhead}, the wedge product on the LHS of \eqref{EqnSd1} is simply the wedge product of $dM$, \textit{i.e.} of its independent elements, hence using the wedge product on the RHS for a real symmetric matrix one gets
\eqna{
(dM)&\equiv\bigwedge_{1\leq i\leq j\leq N}dM_{ij}\\
&=\bigwedge_{1\leq i\leq N}d\lambda_i\bigwedge_{1\leq i<j\leq N}(\lambda_j-\lambda_i)(U^\dagger dU)_{ij}\\
&=\prod_{1\leq i<j\leq N}(\lambda_j-\lambda_i)\bigwedge_{1\leq i\leq N}d\lambda_i\bigwedge_{1\leq i<j\leq N}(U^\dagger dU)_{ij},
}
while for a complex symmetric matrix one obtains
\eqna{
(dM)&\equiv\bigwedge_{1\leq i\leq j\leq N}d\text{Re}{M}_{ij}\wedge d\text{Im}{M}_{ij}\\
&=\bigwedge_{1\leq i\leq N}d\lambda_i\wedge2\lambda_i\text{Im}(U^\dagger dU)_{ii}\bigwedge_{1\leq i<j\leq N}(\lambda_j-\lambda_i)\text{Re}(U^\dagger dU)_{ij}\wedge(\lambda_j+\lambda_i)\text{Im}(U^\dagger dU)_{ij}\\
&=2^N\prod_{1\leq i\leq N}\lambda_i\prod_{1\leq i<j\leq N}(\lambda_j^2-\lambda_i^2)\bigwedge_{1\leq i\leq N}d\lambda_i\bigwedge_{1\leq i\leq j\leq N}(U^\dagger dU)_{ij}.
}
Hence, the measure is
\eqn{dM=c_{SN}^\beta\prod_{1\leq i<j\leq N}|\lambda_i^\beta-\lambda_j^\beta|\prod_{1\leq i\leq N}\lambda_i^{\beta-1}d\lambda_iU^\dagger dU,}[EqnSmeasure]
where the normalization constant is
\eqn{c_{SN}^\beta=\frac{2^{(\beta-1)N}}{2^NN!}=\frac{1}{2^{(2-\beta)N}N!},}[EqnSconstant]
and the integration region is extended to $-\infty<\lambda_i<\infty$ ($0<\lambda_i<\infty$) for all eigen (singular) values [hence the factor of $N!$ in the denominator of \eqref{EqnSconstant}].  The extra factor of $2^N$ in the denominator of \eqref{EqnSconstant} accounts for the remaining freedom in the eigen (singular) value decomposition $M=UDU^T$ where $U$ is replaced by $US$ with $S=\text{diag}(\pm1,\cdots,\pm1)$.  This factor implies the integration region for the group measure is the full region with the Stiefel volume mentioned above.

\subsubsection{Real and Complex Arbitrary Matrix}

The proof of the measure for an arbitrary (denoted by the subscript $A$) matrix is mostly equivalent.  For a real (complex) $N\times N$ arbitrary matrix $M$, the singular value decomposition theorem implies that there exist real (complex) matrices $U$ and $V$ satisfying $U^\dagger U=\mathds{1}$ and $V^\dagger V=\mathds{1}$ such that $M=UDV^\dagger$ where $D$ is real diagonal, \textit{i.e.} $D=\text{diag}(\lambda_1,\cdots,\lambda_N)$, with $\lambda_N>\lambda_{N-1}>\cdots>\lambda_1>0$.  Therefore, one has
\eqn{dM=d(UDV^\dagger)=dUDV^\dagger+UdDV^\dagger+UDdV^\dagger.}[EqnAd]
Multiplying \eqref{EqnSd} by $U^\dagger$ on the left and by $V$ on the right leads to
\eqna{
U^\dagger dMV&=(U^\dagger dU)D+dD+D(V^\dagger dV)^\dagger\\
&=(U^\dagger dU)D+dD-D(V^\dagger dV)\\
&=\left\{\begin{array}{lll}d\lambda_i+\lambda_i[(U^\dagger dU)_{ii}-(V^\dagger dV)_{ii}]&\text{for}&i=j\\\lambda_j(U^\dagger dU)_{ij}-\lambda_i(V^\dagger dV)_{ij}&\text{for}&i\neq j\end{array}\right.,
}[EqnAd1]
where the last two equalities come from $V^\dagger dV=d(V^\dagger V)-dV^\dagger V=-(V^\dagger dV)^\dagger$.  Again, the wedge product on the LHS of \eqref{EqnAd1} is simply the wedge product of $dM$.  Hence using the wedge product on the RHS for a real arbitrary matrix one obtains
\eqna{
(dM)&\equiv\bigwedge_{1\leq i,j\leq N}dM_{ij}\\
&=\bigwedge_{1\leq i\leq N}d\lambda_i\bigwedge_{1\leq i\neq j\leq N}[\lambda_j(U^\dagger dU)_{ij}-\lambda_i(V^\dagger dV)_{ij}]\\
&=\bigwedge_{1\leq i\leq N}d\lambda_i\bigwedge_{1\leq i<j\leq N}[\lambda_j(U^\dagger dU)_{ij}-\lambda_i(V^\dagger dV)_{ij}]\wedge[\lambda_j(V^\dagger dV)_{ij}-\lambda_i(U^\dagger dU)_{ij}]\\
&=\prod_{1\leq i<j\leq N}(\lambda_j^2-\lambda_i^2)\bigwedge_{1\leq i\leq N}d\lambda_i\bigwedge_{1\leq i<j\leq N}(U^\dagger dU)_{ij}\wedge(V^\dagger dV)_{ij},
}
while for a complex arbitrary matrix one gets
\small
\eqna{
(dM)&\equiv\bigwedge_{1\leq i,j\leq N}d\text{Re}{M}_{ij}\wedge d\text{Im}{M}_{ij}\\
&=\bigwedge_{1\leq i\leq N}d\lambda_i\wedge\lambda_i\text{Im}[(U^\dagger dU)_{ii}-(V^\dagger dV)_{ii}]\\
&\phantom{=}\hspace{0.5cm}\bigwedge_{1\leq i\neq j\leq N}[\lambda_j\text{Re}(U^\dagger dU)_{ij}-\lambda_i\text{Re}(V^\dagger dV)_{ij}]\wedge[\lambda_j\text{Im}(U^\dagger dU)_{ij}-\lambda_i\text{Im}(V^\dagger dV)_{ij}]\\
&=\bigwedge_{1\leq i\leq N}d\lambda_i\wedge\lambda_i\text{Im}[(U^\dagger dU)_{ii}-(V^\dagger dV)_{ii}]\\
&\phantom{=}\hspace{0.5cm}\bigwedge_{1\leq i<j\leq N}[\lambda_j\text{Re}(U^\dagger dU)_{ij}-\lambda_i\text{Re}(V^\dagger dV)_{ij}]\wedge[\lambda_j\text{Re}(V^\dagger dV)_{ij}-\lambda_i\text{Re}(U^\dagger dU)_{ij}]\\
&\phantom{=}\hspace{1cm}\bigwedge_{1\leq i<j\leq N}[\lambda_j\text{Im}(U^\dagger dU)_{ij}-\lambda_i\text{Im}(V^\dagger dV)_{ij}]\wedge[\lambda_i\text{Im}(U^\dagger dU)_{ij}-\lambda_j\text{Im}(V^\dagger dV)_{ij}]\\
&=\prod_{1\leq i\leq N}\lambda_i\prod_{1\leq i<j\leq N}(\lambda_j^2-\lambda_i^2)^2\bigwedge_{1\leq i\leq N}d\lambda_i\wedge\text{Im}[(U^\dagger dU)_{ii}-(V^\dagger dV)_{ii}]\bigwedge_{1\leq i<j\leq N}(U^\dagger dU)_{ij}\wedge(V^\dagger dV)_{ij}.
}
\normalsize
Therefore, the final measure becomes
\eqn{dM=c_{AN}^\beta\prod_{1\leq i<j\leq N}|\lambda_i^2-\lambda_j^2|^\beta\prod_{1\leq i\leq N}\lambda_i^{\beta-1}d\lambda_i(U^\dagger dU)'(V^\dagger dV)',}[EqnAmeasure]
where the normalization constant is
\eqn{c_{AN}^\beta=\frac{1}{2^{(2-\beta)N}N!},}[EqnAconstant]
and the integration region is over all non-negative singular values, \textit{i.e.} $0<\lambda_i<\infty$.  Again, this accounts for the factor of $N!$ in the denominator of \eqref{EqnAconstant}.  The other factor of $2^{(2-\beta)N}$ in the denominator of \eqref{EqnAconstant} accounts for the remaining freedom in the singular value decomposition $M=UDV^\dagger$ where $U$ and $V$ are replaced respectively by $US$ and $VS$ with $S=\text{diag}(\pm1,\cdots,\pm1)$.  For a real arbitrary matrix, this factor implies the group measure is integrated over the full region with the Stiefel volume mentioned above.  For a complex arbitrary matrix, the remaining freedom is already taken care of by the fact that not all diagonal elements of $U^\dagger dU$ and $V^\dagger dV$ appear in the measure \eqref{EqnAmeasure} but only the specific combination $\text{Im}[(U^\dagger dU)_{ii}-(V^\dagger dV)_{ii}]$ does.  This observation is denoted by a prime on $U^\dagger dU$ and $V^\dagger dV$.  Although it is counter-intuitive to have more integration parameters on the RHS than the LHS of \eqref{EqnAmeasure}, it is always possible to introduce the missing integration variables [for example, $\text{Im}(U^\dagger dU)_{ii}$] and compensate by dividing the measure by the appropriate volume, \textit{i.e.} $(2\pi)^{(\beta-1)N}$.  This trick allows integrating over the full Stiefel volumes of both $U$ and $V$ in the complex case also, as in subsection \ref{SsDM}.


\section{Circular Law}\label{SCircle}

In this appendix the different circular laws for Gaussian ensembles are re-derived from the Coulomb gas approach.

\subsection{Level Density}

The level density $\rho(x)$ for real ($\beta=1$) and complex ($\beta=2$), symmetric ($S$) and arbitrary ($A$), matrices $M$ with usual Gaussian ensembles for the singular values $\lambda_i$ given by
\eqna{
P_S(\lambda)d\lambda&\propto\prod_{1\leq i<j\leq N}|\lambda_i^\beta-\lambda_j^\beta|\prod_{1\leq i\leq N}|\lambda_i|^{\beta-1}e^{-\lambda_i^2}d\lambda_i,\\
P_A(\lambda)d\lambda&\propto\prod_{1\leq i<j\leq N}|\lambda_i^2-\lambda_j^2|^\beta\prod_{1\leq i\leq N}\lambda_i^{\beta-1}e^{-\lambda_i^2}d\lambda_i,
}
[compare to \eqref{EqnPDFdmh}] can be computed from the associated Coulomb-like gas with Hamiltonians
\eqna{
H_S&=-\frac{1}{2}\int dxdy\,\rho_{SN}^\beta(x)\rho_{SN}^\beta(y)\ln|x^\beta-y^\beta|-(\beta-1)\int dx\,\rho_{SN}^\beta(x)\ln|x|+\int dx\,\rho_{SN}^\beta(x)x^2,\\
H_A&=-\frac{1}{2}\int dxdy\,\rho_{AN}^\beta(x)\rho_{AN}^\beta(y)\ln|x^2-y^2|-\left(1-\frac{1}{\beta}\right)\int dx\,\rho_{AN}^\beta(x)\ln(x)+\frac{1}{\beta}\int dx\,\rho_{AN}^\beta(x)x^2.
}[EqnHSA]
Extremizing \eqref{EqnHSA} with respect to the level density, subject to the normalization constraint $\int dx\,\rho(x)=N$, leads to
\eqna{
\xi_S&=-\int dy\,\rho_{SN}^\beta(y)\ln|x^\beta-y^\beta|-(\beta-1)\ln|x|+x^2,\\
\xi_A&=-\int dy\,\rho_{AN}^\beta(y)\ln|x^2-y^2|-\left(1-\frac{1}{\beta}\right)\ln(x)+\frac{1}{\beta}x^2,
}[EqnrhoSA]
where $\xi$ is the constant Lagrange multiplier.  Deriving \eqref{EqnrhoSA} with respect to $x$ gives
\eqna{
0&=-\beta x^{\beta-1}\mathcal{P}\int dy\,\frac{\rho_{SN}^\beta(y)}{x^\beta-y^\beta}-\frac{\beta-1}{x}+2x,\\
0&=-2x\mathcal{P}\int dy\,\frac{\rho_{AN}^\beta(y)}{x^2-y^2}-\left(1-\frac{1}{\beta}\right)\frac{1}{x}+\frac{2}{\beta}x,
}[EqndrhoSA]
where $\mathcal{P}$ denotes the principal value.  At this point, it is convenient to rescale the variables such that the thermodynamic limit with $N$ large can be easily taken.  This is done with the help of $x=\sqrt{N}\hat{x}$ and $\rho(x)=\sqrt{N}\hat{\rho}(\hat{x})$ which change \eqref{EqndrhoSA} to
\eqna{
0&=-\beta\hat{x}^{\beta-1}\mathcal{P}\int d\hat{y}\,\frac{\hat{\rho}_{SN}^\beta(\hat{y})}{\hat{x}^\beta-\hat{y}^\beta}-\frac{\beta-1}{N\hat{x}}+2\hat{x},\\
0&=-2\hat{x}\mathcal{P}\int d\hat{y}\,\frac{\hat{\rho}_{AN}^\beta(\hat{y})}{\hat{x}^2-\hat{y}^2}-\left(1-\frac{1}{\beta}\right)\frac{1}{N\hat{x}}+\frac{2}{\beta}\hat{x},
}[EqndrhohSA]
and the normalization condition to $\int d\hat{x}\,\hat{\rho}(\hat{x})=1$.  In the thermodynamic limit, the second term in each equation of \eqref{EqndrhohSA} is negligible and can be discarded.  From now on the normalized level density is easily obtained from the usual treatment and is given by
\eqna{
\hat{\rho}_{SN}^\beta(\hat{x})&=\frac{2}{\pi}\sqrt{\beta-\hat{x}^2}\quad\quad(\beta-2)\sqrt{\beta}\leq\hat{x}\leq\sqrt{\beta},\\
\hat{\rho}_{AN}^\beta(\hat{x})&=\frac{2}{\beta\pi}\sqrt	{2\beta-\hat{x}^2}\quad\quad0\leq\hat{x}\leq\sqrt{2\beta}.
}[EqnrhoSAsoln]
Note that the small differences between the circular laws obtained here and the usual ones come from the different normalizations used.


\bibliography{NeutrinoRMT}

\end{document}